\documentclass[aps,physrev,preprint,groupedaddress,showkeys]{revtex4-2}

\usepackage{graphicx}% Include figure files
\usepackage{dcolumn}% Align table columns on decimal point
\usepackage{bm}% bold math
\usepackage[colorlinks,linkcolor=blue,anchorcolor=blue,citecolor=blue]{hyperref}
\usepackage{tabularx}
\usepackage[mathlines]{lineno}
\newcommand{\pip}{\pi^+}
\newcommand{\pim}{\pi^-}
\newcommand{\piz}{\pi^0}

\newcommand{\azz}{a^{0}_{0}(980)}
\newcommand{\fz}{f_{0}(980)}

\begin{document}

%\preprint{}

\title{The width of $f_{0}(980)$ in isospin-symmetry-breaking decays}


% \affiliation command should follow the other information
\author{Xiaolong Du$^1$}  

\author{Yun Liang$^1$}

\author{Wencheng Yan$^1$}
\email[]{yanwc@zzu.edu.cn}

\author{Demin Li$^1$}
\email[]{lidm@zzu.edu.cn}

%\thanks{}
\affiliation{School of Physics and Microelectronics, Zhengzhou University, Zhengzhou 450001, P.R.China}


\date{\today}

\begin{abstract}
The scalar meson $f_{0}(980)$ has long posed a perplexing puzzle within the realm of light hadron physics. Conventionally, its mass and width in normal decay processes have been estimated as $M=990\pm20$~MeV/$c^2$ and $\Gamma=40-100$~MeV, respectively. Theoretical explanations regarding the internal structure of $f_{0}(980)$ range from it being a conventional quark-antiquark meson to a tetraquark state, a $K\overline{K}$ molecule, or even a quark-antiquark gluon hybrid. However, a definitive consensus has remained elusive over a considerable duration. Recent observations by the BESIII experiment have unveiled anomalously narrow widths of $f_{0}(980)$ in five independent isospin-symmetry-breaking decay channels. Harnessing these experimental findings, we performed a simultaneous fit to the $\pi\pi$ invariant mass distributions, resulting in a refined determination of the mass and width in isospin-symmetry-breaking decays as $M=990.0\pm0.4(\text{stat})\pm0.1(\text{syst})$~MeV/$c^2$ and $\Gamma=11.4\pm1.1(\text{stat})\pm0.9(\text{syst})$~MeV, respectively. Here, the first errors are statistical and the second are systematic. Furthermore, by employing the parameterized Flatt\'{e} formula to fit the same $\pi\pi$ invariant mass distributions, we ascertained the values of the two coupling constants, $g_{f\pi\pi}$ and $g_{fK\overline{K}}$, as $g_{f\pi\pi}=0.46\pm0.03$ and $g_{fK\overline{K}}=1.24\pm0.32$, respectively. Based on the joint confidence regions of $g_{f\pi\pi}$ and $g_{fK\overline{K}}$, we draw the conclusion that the experimental data exhibit a propensity to favor the $K\overline{K}$ molecule model and the quark-antiquark ($q\bar{q}$) model, while offering relatively less support for the tetraquarks ($q^{2}\bar{q}^{2}$) model and the quark-antiquark gluon ($q\bar{q}g$) hybrid model.
\end{abstract}


\maketitle

%\section{}
%\subsection{}
%\subsubsection{}

%%%%%%%%%%%%%%%%%%%%%%%%%%%%%%%%%%%%%%%%%%
\section{Introduction \label{intro}}

The majority of hadrons manifest as resonances, appearing as poles of the S-matrix on unphysical sheets in the complex plane. The mass and width of a resonance correspond to the location of the nearest pole on an unphysical sheet of the complex energy plane, commonly denoted as $\sqrt{s_{pole}}=M-i\Gamma/2$. Alongside the pole position, another important characteristic of a resonance is its coupling to different channels, which can be described by the pole residues. The pole position and pole residues are theoretically independent of any specific model and serve as reliable descriptions for a resonance~\cite{PDG2022}. When a resonance is narrow and if there is no relevant thresholds or other resonances nearby, its properties can be described using the Breit–Wigner parameterization.

The scalar meson $f_{0}(980)$, characterized by its quantum numbers $I^{G}J^{PC}=0^{+}0^{++}$, was experimentally established four decades ago. However, its nature has remained a long-standing puzzle. This scalar meson primarily decays into final states involving $\pi\pi$, while also exhibiting a small branching fraction for $K\bar{K}$ decays~\cite{PDG2022}. The resonant parameters of $f_{0}(980)$ have been determined through various decay channels. For instance, measurements of decay processes such as $\phi\to\pi^{0}\pi^{0}\gamma$ by SND~\cite{SND2000} and KLOE~\cite{KLOE2000}, and $J/\psi \to \phi \pi^{+} \pi^{-}$ by BESII~\cite{BESII2005}, as well as analyses of scattering processes like $e^{+}e^{-} \to K^{+}K^{-}\pi^{+}\pi^{-} / K^{+}K^{-}\pi^{0}\pi^{0}$ by BABAR~\cite{BABAR2007}, $\pi\pi$ scattering data, and $K_{l4}$ decay data~\cite{Martin2011PRD,Martin2011PRL}, have provided insights into the mass and width of $f_{0}(980)$. The Particle Data Group (PDG) estimated the mass to be $M=990\pm20$~MeV/$c^2$ and the width to range from $\Gamma=40$ to $100$~MeV until 2016~\cite{PDG2014}. It is important to note that the shape of $f_{0}(980)$ varies depending on the specific decay process, indicating that not all uncertainties in the parameters stem from experimental data.

In 2012, the BESIII Collaboration made the initial observation of anomalously narrow widths of approximately 10~MeV for $f_{0}(980)$ through isospin-symmetry-breaking decays of $J/\psi\to\gamma\eta(1405), \eta(1405)\to\pi^{0} f_{0}(980) \to \pi^{0}\pi^{+}\pi^{-}/ \pi^{0}\pi^{0}\pi^{0}$~\cite{WuZ2012}. The isospin-symmetry-breaking ratio between $\eta(1405)\to\pi^{0} f_{0}(980) \to \pi^{0}\pi^{+}\pi^{-}$ and $\eta(1405)\to \pi^{0} a^{0}_{0}(980) \to \pi^{0}\eta\pi^{0}$ was measured to be as high as $(17.9\pm0.42)\%$. Three years later, the BESIII Collaboration once again observed similarly narrow widths for $f_{0}(980)$ in isospin-symmetry-breaking decays of $J/\psi\to\phi f_{1}(1285), f_{1}(1285)\to\pi^{0}\fz \to\pi^{0}\pi^{+}\pi^{-} /\pi^{0}\pi^{0}\pi^{0}$, with a measured isospin-symmetry-breaking ratio of $(3.6\pm1.4)\%$~\cite{XiaLG2015}. Subsequently, the PDG updated the width of $f_{0}(980)$ to be $\Gamma=$10 to $100$~MeV~\cite{PDG2022}. In 2018, the BESIII Collaboration further confirmed the anomalously narrow width of $f_{0}(980)$ through an isospin-symmetry-breaking decay of $\chi_{c1}\to \pi^{0} f_{0}(980)\to\pi^{0}\pi^{+}\pi^{-}$~\cite{YanWC2018}. It is noteworthy that these narrow-width $\fz$ mesons are solely produced in isospin-symmetry-breaking decays, which starkly contrasts with the normal width of $f_{0}(980)$ observed in isospin-symmetry-conserving decays.

\begin{figure}[htbp]
\centering
\mbox{
\put(10,70){\textbf{(a)}}
\includegraphics[width=5.2cm]{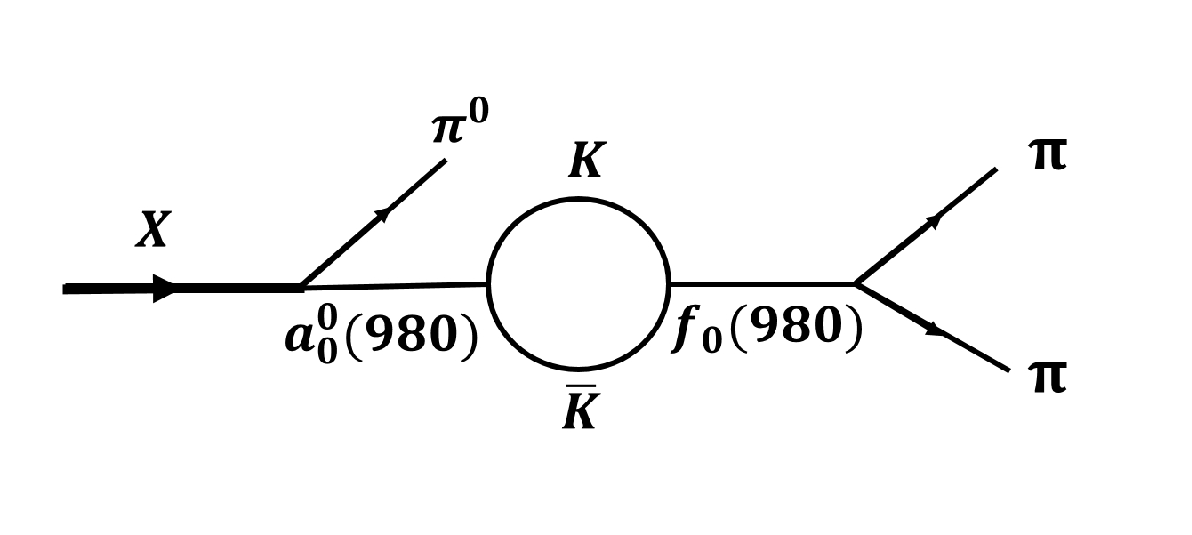}
\put(10,70){\textbf{(b)}}
\includegraphics[width=5.2cm]{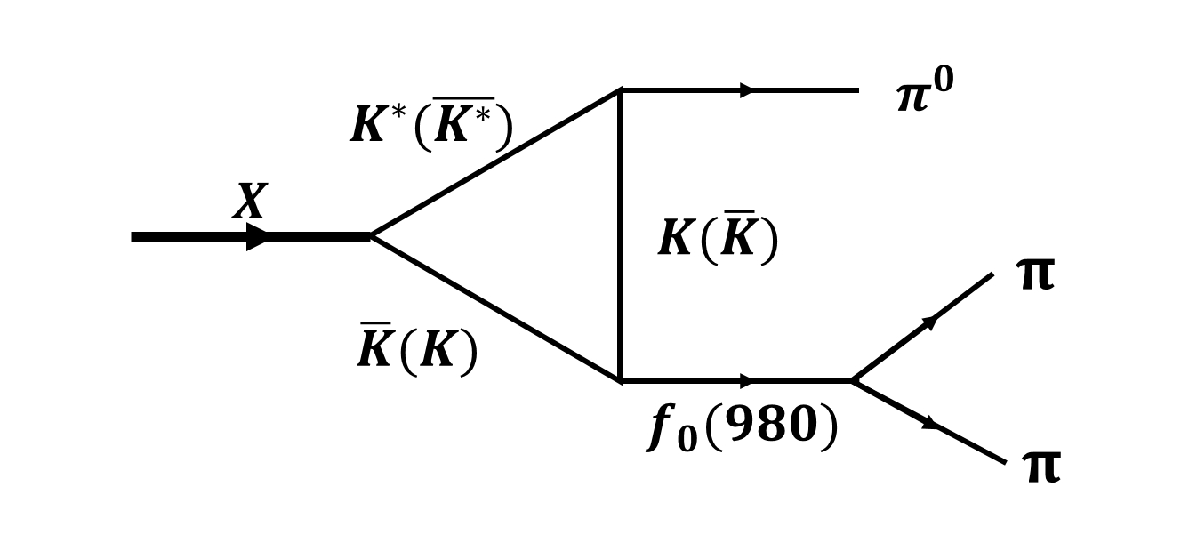}}
\caption{The Feynman diagram of the hadronic level: (\textbf{a}) The diagram of $a^{0}_{0}(980) \to f_{0}(980)$ mixing in the decay of $X \to \pi^{0} a^{0}_{0}(980) \to \pi^{0} f_{0}(980) \to 3\pi$. (\textbf{b}) The diagram of triangle singularity mechanism in the decay of $X \to \pi^{0} f_{0}(980) \to 3\pi$. Here, $X$ in the two diagrams can be $\eta(1405)$, $f_{1}(1285)$ or $\chi_{c1}$.
\label{fig1}}
\end{figure}

Theoretically, the internal structure of the $f_{0}(980)$ is not only considered as the conventional quark-antiquark~\cite{Achasov1989}, but also proposed to be tetraquarks~\cite{Achasov1989,qqqq1983}, $K\overline{K}$ molecule~\cite{KK1990}, or quark-antiquark gluon hybrid~\cite{qqg}. However, explanations about the nature of $f_{0}(980)$ have been controversial to date. The most renowned theoretical study about $f_{0}(980)$ is the $a^{0}_{0}(980)$-$f_{0}(980)$ mixing mechanism, which was first proposed in the late 1970s~\cite{Achasov1979}. Since both $a^{0}_{0}(980)$ and $f_{0}(980)$ can decay into $K\overline{K}$, the charged and neutral kaon mass thresholds differ by approximately 8~MeV due to isospin-symmetry-breaking effects. The mixing amplitude between $a^{0}_{0}(980)$ and $f_{0}(980)$ is predominantly influenced by unitary cuts of the intermediate two-kaon system and proportional to the phase-space difference between the charged and neutral kaon systems. As a consequence, a narrow peak of about 8~MeV in width is predicted between the charged and neutral $K\bar{K}$ mass thresholds. The Feynman diagram of $a^{0}_{0}(980)\to f_{0}(980)$ mixing in the decay of $X \to \pi^{0} a^{0}_{0}(980) \to \pi^{0} f_{0}(980) \to 3\pi$ is shown in Figure~\ref{fig1}(a), where $X$ can be $\eta(1405)$, $f_{1}(1285)$ or $\chi_{c1}$. The $a^{0}_{0}(980)$-$f_{0}(980)$ mixing mechanism has been investigated extensively for a long time, and many decay processes have been discussed~\cite{Kerbikov, Achasov20041, Achasov20042, Kudr1, Kudr2, Kudr3, Grishina, Close1, Close2, Hanhart, WuJJ2007, WuJJ2008}. There was no experimental results until the BESIII Collaboration reported $a^{0}_{0}(980)$-$f_{0}(980)$ mixing via the decays of $J/\psi \to\phi\fz \to\phi\azz \to\phi\eta\piz$ and $\chi_{c1} \to \piz\azz \to\piz\fz \to\piz\pip\pim$~\cite{WangYD2011, YanWC2018}. The mixing intensity of $a^{0}_{0}(980) \to f_{0}(980)$, \emph{i.e.} the isospin-symmetry-breaking ratio, is measure to be $0.40\pm0.07\pm0.14\pm0.07$, which is less than $1.0\%$. Obviously, the $a^{0}_{0}(980)$-$f_{0}(980)$ mixing mechanism can not completely describe the large isospin-symmetry-breaking ratio in the decays of $\eta(1405) \to \pi^{0} f_{0}(980) \to \pi^{0}\pi^{+}\pi^{-}/\pi^{0}\pi^{0}\pi^{0}$ and $f_{1}(1285) \to \pi^{0} f_{0}(980) \to \pi^{0}\pi^{+}\pi^{-}/\pi^{0}\pi^{0}\pi^{0}$.

\unskip

Since the BESIII experiment reported the anomalously narrow widths of approximately 10~MeV of $f_{0}(980)$ mesons~\cite{WuZ2012}, a novel explanation known as the triangle singularity mechanism was proposed to elucidate its internal behavior in isospin-symmetry-breaking decay processes~\cite{WuJJ2012}. Taking the decay $\eta(1405)\to\pi^{0} f_{0}(980) \to 3\pi$ as an example, the intermediate $K\bar{K}^{*}+$c.c. pair can exchange an on-shell kaon, satisfying the energy-momentum conservation at the three interaction vertices. Consequently, the physical amplitude exhibits a logarithmic triangle singularity, and the kinematic effects lead to a narrow peak in the $\pi\pi$ invariant mass distributions. Figure~\ref{fig1}(b) depicts the Feynman diagram representing the triangle singularity mechanism. This mechanism can effectively account for the narrow width of $\fz$ mesons and the significant isospin-symmetry-breaking ratio observed in the decays of $\eta(1405) \to \pi^{0}\fz \to 3\pi$. Further in-depth discussions and extensive applications of the triangle singularity mechanism have been conducted~\cite{Aceti2012, DuMC2019, DuMC2022}, encompassing explaining other decay processes~\cite{Wang2017, JingHJ2019, SaKai2020, Natsumi2021, Feijoo20}.

This article investigates the width of the $f_{0}(980)$ meson in the context of isospin-symmetry-breaking decays. Firstly, we employ the non-relativistic Breit-Wigner formula to perform a simultaneous fit to the five $\pi\pi$ invariant mass distributions reported by the BESIII Collaboration. This enables us to accurately determine the mass and width of the $f_{0}(980)$ meson in isospin-symmetry-breaking decays, as discussed in Section~\ref{simfit}. Secondly, we adopt the energy-dependent Flatt\'{e} formula as a shape description for $f_{0}(980)$ and conduct a simultaneous fit to the $\pi\pi$ invariant mass distributions. This allows us to extract the coupling constants of $f_{0}(980)\to K\overline{K}$ and $f_{0}(980)\to\pi\pi$, denoted as $g_{fK\bar{K}}$ and $g_{f\pi\pi}$, respectively, as elaborated in Section~\ref{coupconst}. Thirdly, we establish a joint confidence region for the two coupling constants, providing quantitative constraints on different theoretical models of the $f_{0}(980)$, as discussed in Section~\ref{confiregion}. Finally, in Section~\ref{summ}, we summarize the key findings and conclusions of this article.

%%%%%%%%%%%%%%%%%%%%%%%%%%%%%%%%%%%%%%%%%%
\section{Simultaneous fit to $\pi\pi$ invariant mass distributions \label{simfit}}

The BESIII Collaboration has reported the presence of an anomalously narrow width for the $f_{0}(980)$ meson in five distinct isospin-symmetry-breaking decay processes. Specifically, these processes include $J/\psi \to\gamma\eta(1405), \eta(1405)\to\pi^{0} f_{0}(980) \to \pi^{0}\pi^{+}\pi^{-}/\pi^{0}\pi^{0}\pi^{0}$~\cite{WuZ2012}, $J/\psi\to\phi f_{1}(1285), f_{1}(1285)\to\pi^{0} f_{0}(980) \to \pi^{0}\pi^{+}\pi^{-}/\pi^{0}\pi^{0}\pi^{0}$~\cite{XiaLG2015}, and $\psi(2S) \to \gamma \chi_{c1}, \chi_{c1} \to \pi^{0} f_{0}(980) \to \pi^{0}\pi^{+}\pi^{-}$~\cite{YanWC2018}. We acquire the binned data, presented as histogram data, from the aforementioned published results. In order to accurately determine the mass and width of the $f_{0}(980)$ meson in isospin-symmetry-breaking decays, we perform a simultaneous fit to the $\pi\pi$ invariant mass spectra of the five decay channels mentioned above, utilizing the RooFit toolkit software~\cite{roofit}.

Because the width of the $f_{0}(980)$ meson is narrow enough here, the line shape of $f_{0}(980)$ can be parameterized by a non-relativistic Breit-Wigner formula:
\begin{linenomath}
\begin{equation}
BW_{f} = \frac{\Gamma_{el}/2}{m_{f} - \sqrt{s} - i\Gamma_{f}/2},
\label{breitwigner}
\end{equation}
\end{linenomath}
where, $\Gamma_{el}$ is called elastic width, which is a constant, $m_{f}$ is the mean value of $f_{0}(980)$ mass, $\sqrt{s}$ is the center-of-mass energy, and $\Gamma_{f}$ is the total width of the $f_{0}(980)$.

In the simultaneous fit, the determination of the width of the $f_{0}(980)$ meson is significantly influenced by the detector resolutions. Therefore, the detector resolutions are obtained in advance using Monte Carlo simulations and fixed during the fit, listed in Table~\ref{tab1}. The shape of the narrow $f_{0}(980)$ meson is described by the non-relativistic Breit-Wigner function convolved with the corresponding Gaussian mass resolution for each decay channel. The background in each decay channel is modeled by either a first-order or a second-order Chebyshev polynomial, with the polynomial order matching the fit employed in the published papers by BESIII~\cite{WuZ2012, XiaLG2015, YanWC2018}.

We construct the probability density function (pdf) for each decay mode as follows:
\begin{linenomath}
\begin{equation}
\text{pdf} = N_{f} \times (|BW_{f}(m_{f},\Gamma_{f})|^{2}\otimes\text{Gauss}(m_{0},\sigma_{0})) + N_{\text{bkg}} \times \text{Chebyshev-poly.},
\label{pdf}
\end{equation}
\end{linenomath}
where, $BW_{f}(m_{f},\Gamma_{f})$ is the Breit-Wigner formula in Equation~\ref{breitwigner}, $\text{Gauss}(m_{0},\sigma_{0})$ is the Gaussian function accounting for the mass resolution, $\text{Chebyshev-poly.}$ represents the background shape, $N_{f}$ and $N_{\text{bkg}}$ are the numbers of the $f_{0}(980)$ meson and continuum backgrounds, respectively. During the simultaneous fit, we utilize the probability density function of each decay channel along with the corresponding histogram data to construct the likelihood function. Subsequently, the likelihood functions of the five decay channels are summed together. Typically, we take the negative natural logarithm of the likelihood value and utilize the RooFit toolkit to determine the minimum value.

\begin{figure}[htbp]
\centering
\includegraphics[width=5.2cm]{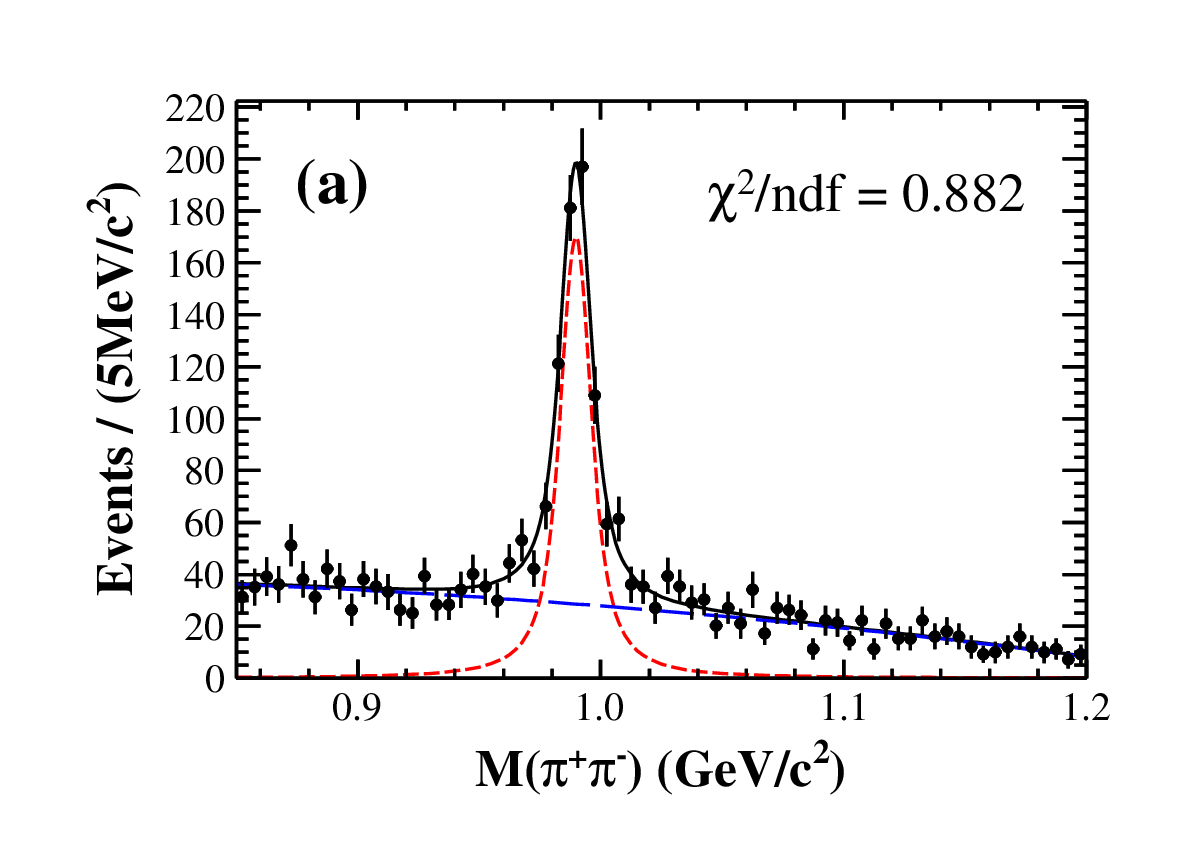}
\includegraphics[width=5.2cm]{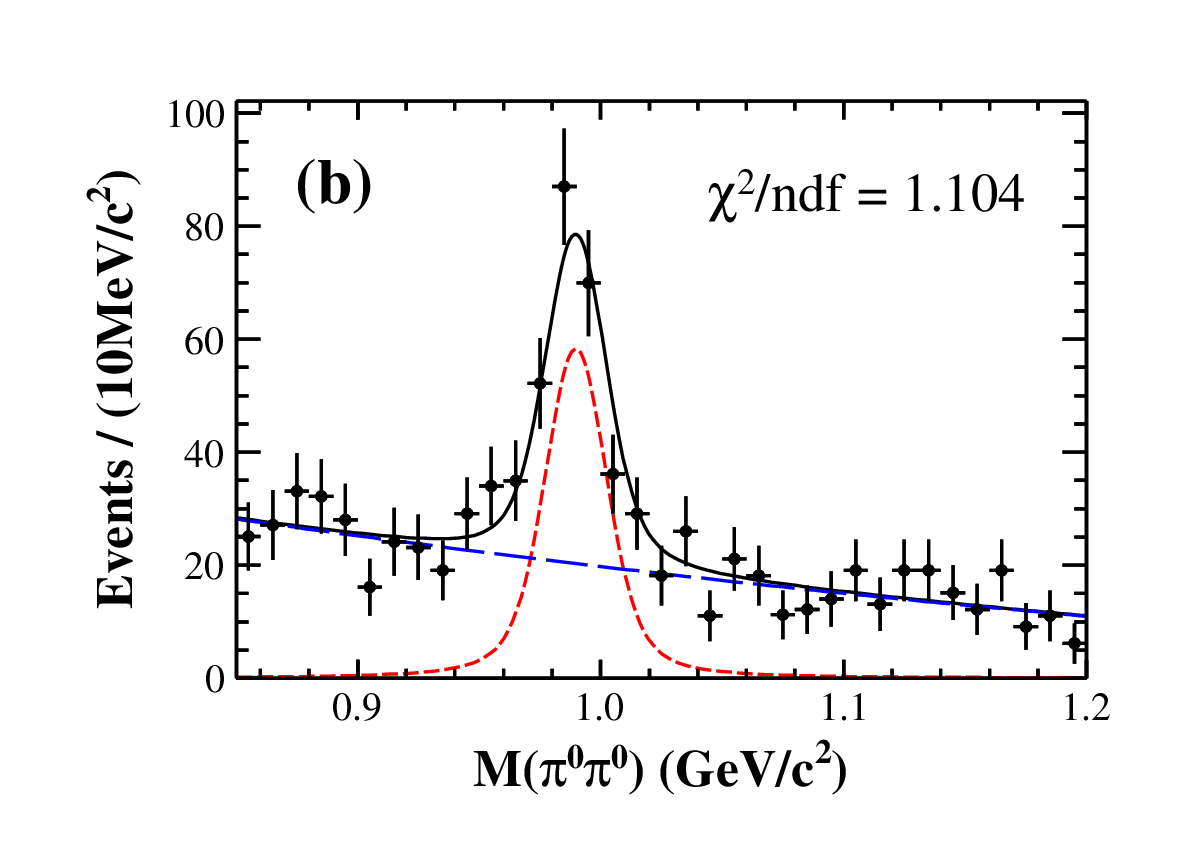}
\includegraphics[width=5.2cm]{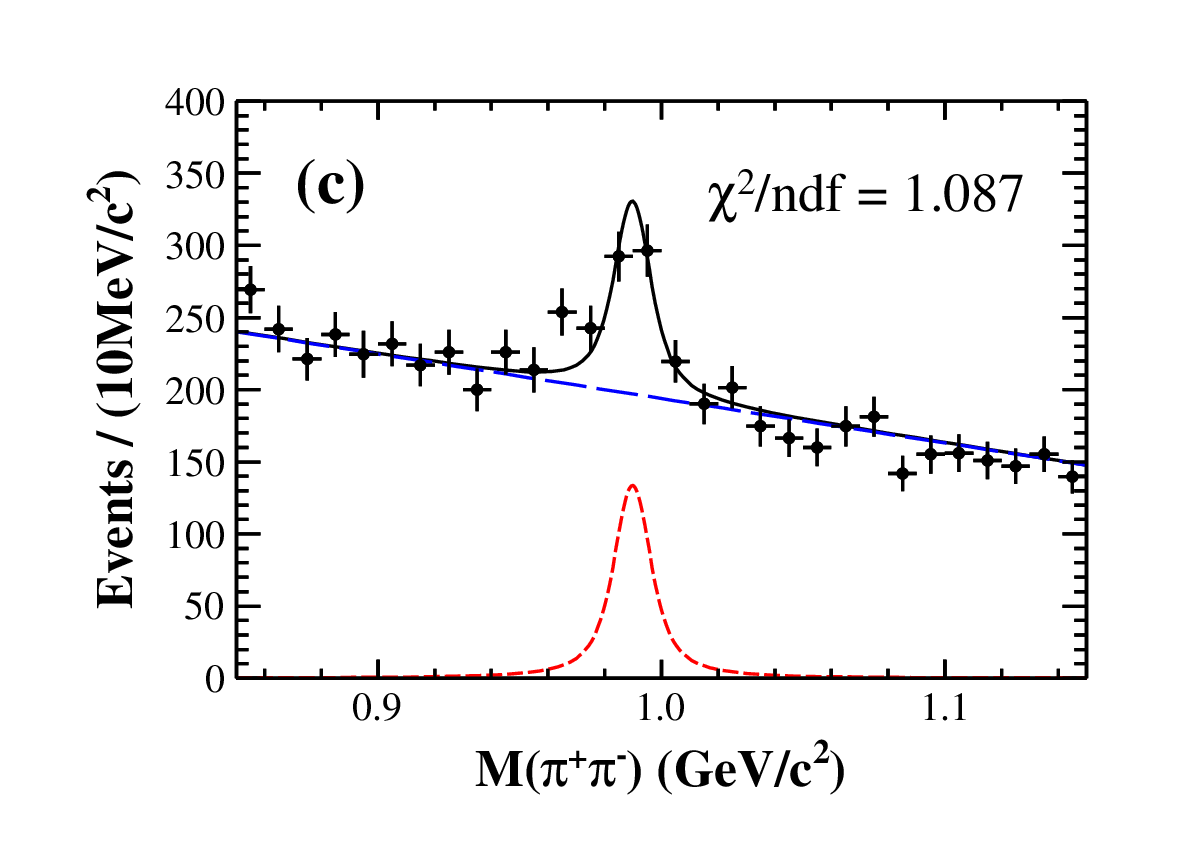}
\includegraphics[width=5.2cm]{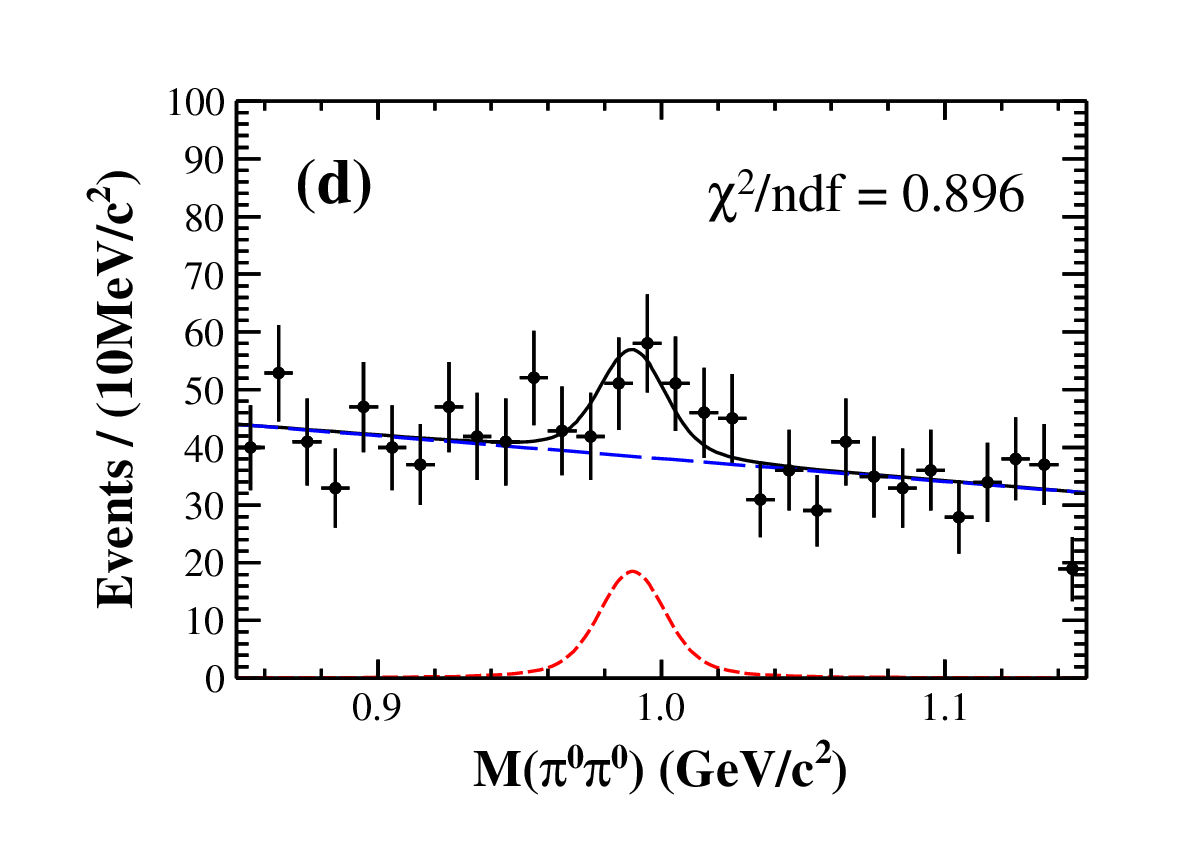}
\includegraphics[width=5.2cm]{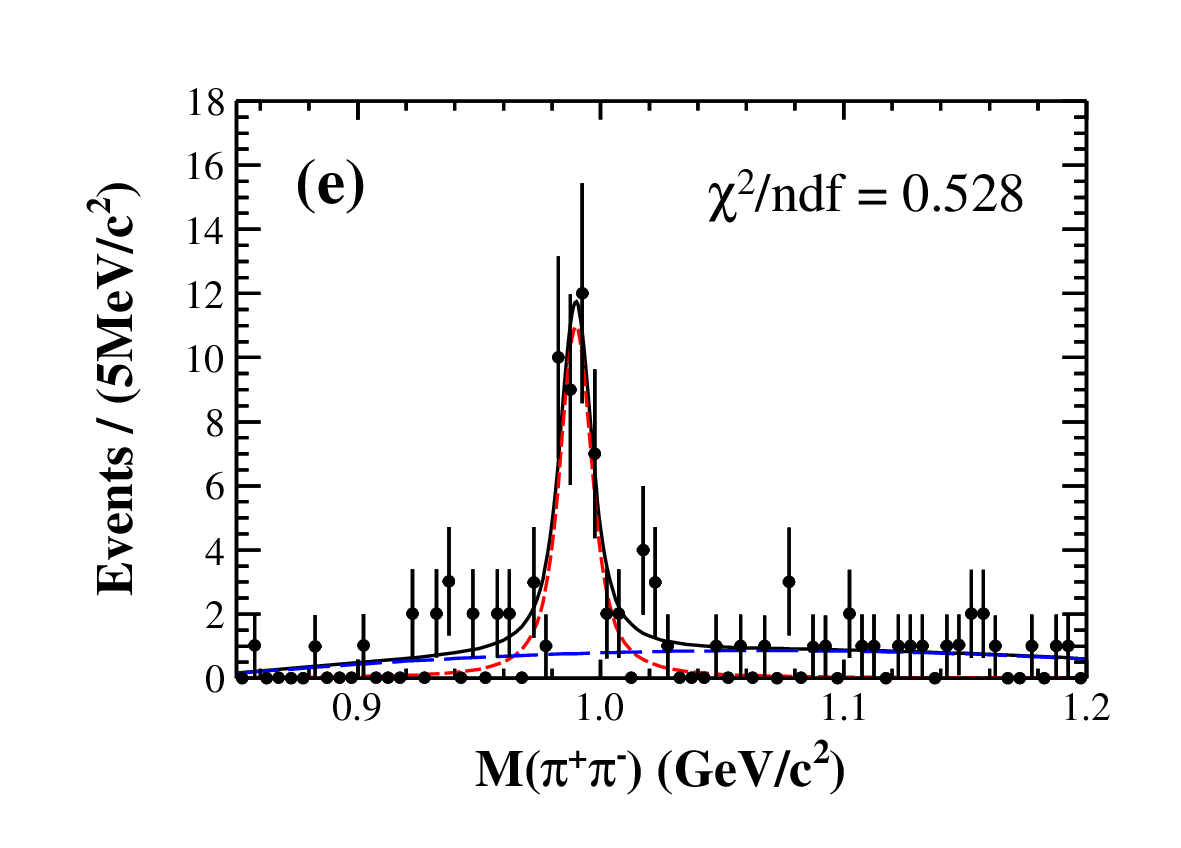}
\includegraphics[width=5.2cm]{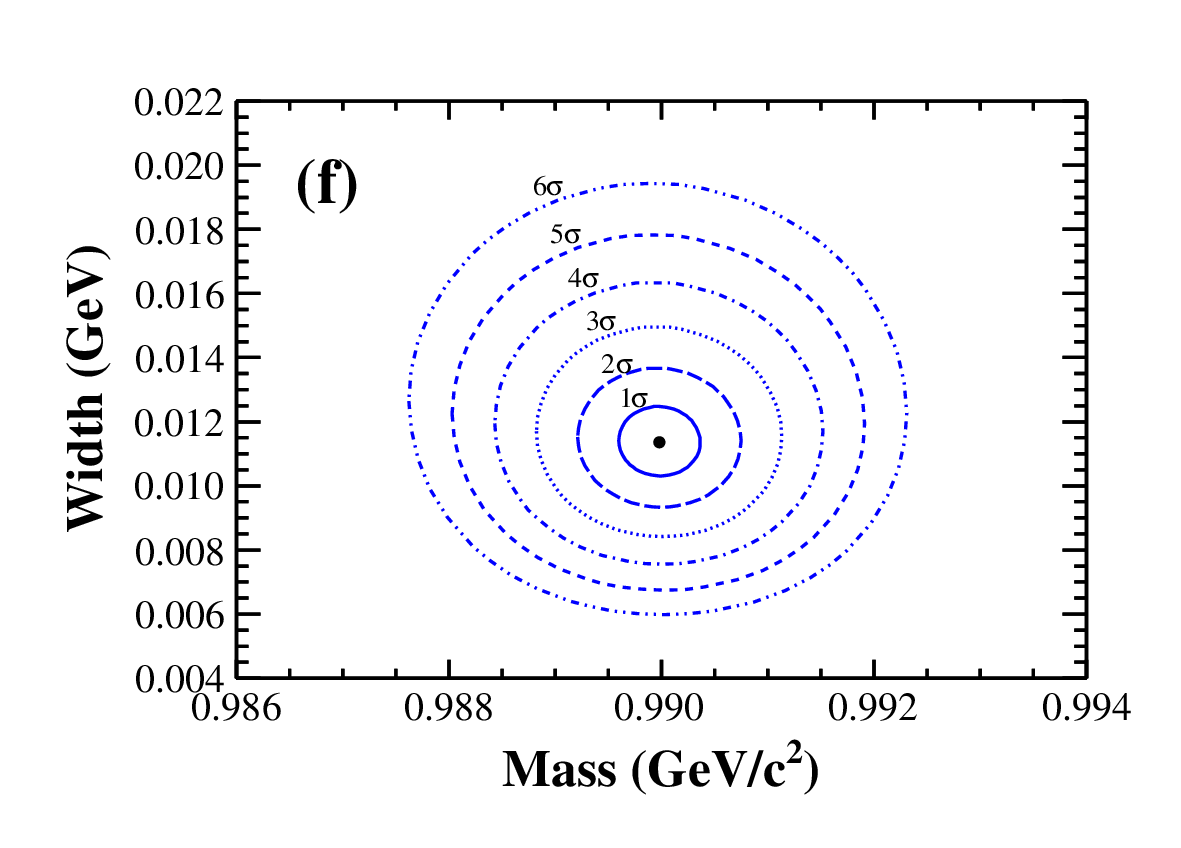}
\caption{The simultaneous fit to the $\pi\pi$ invariant mass spectra of the decay channels of (\textbf{a}) $\eta(1405) \to \pi^{0} \fz \to \pi^{0}\pi^{+}\pi^{-}$, (\textbf{b}) $\eta(1405) \to \pi^{0} \fz \to \pi^{0}\pi^{0}\pi^{0}$; (\textbf{c}) $f_{1}(1285)\to \pi^{0} \fz \to \pi^{0}\pi^{+}\pi^{-}$,  (\textbf{d}) $f_{1}(1285)\to \pi^{0} \fz \to \pi^{0}\pi^{0}\pi^{0}$;  (\textbf{e}) $\chi_{c1} \to \pi^{0} \fz \to \pi^{0}\pi^{+}\pi^{-}$. The dots with error bars are the data reported by BESIII, the solid curves represent the total fit, the dashed curves represent the Breit-Wigner function of the $f_{0}(980)$ meson, and the long-dashed curves represent the background polynomials. (\textbf{f}) The joint confidence regions of the mass and width of the $f_{0}(980)$ meson obtained from the simultaneous fit. \label{fig2}}
\end{figure}
\unskip

The results of the simultaneous fit are illustrated in Figure~\ref{fig2} (a-e). The data obtained from BESIII are depicted as dots with error bars, while the solid curves represent the overall fit results. The dashed curves correspond to the Breit-Wigner function of the $f_{0}(980)$ meson, and the long-dashed curves represent the background polynomials. Additionally, the goodness-of-fit value for each decay channel is indicated on the figures. Figures~\ref{fig2} (a) and (b) display the fit results for $\eta(1405) \to \pi^{0} \fz$, $\fz\to\pi^{+}\pi^{-}$, and $\pi^{0}\pi^{0}$ decays, with corresponding goodness-of-fit values of 0.882 and 1.104, respectively. Figures~\ref{fig2} (c) and (d) exhibit the fit results for $f_{1}(1285) \to \pi^{0} \fz$, $\fz \to\pi^{+}\pi^{-}$, and $\pi^{0}\pi^{0}$ decays, with goodness-of-fit values of 1.087 and 0.896, respectively. Figure~\ref{fig2} (e) shows the fit result for $\chi_{c1} \to \pi^{0} \fz$ and $\fz\to\pi^{+}\pi^{-}$ decay, and the goodness-of-fit value is 0.528, which means the fit a little worse than other channels. 

The mass and width values of the $f_{0}(980)$ meson obtained from the simultaneous fit, along with the results reported by the BESIII Collaboration, are listed in Table~\ref{tab1}. The mass obtained from the fit is $990.0\pm0.4(\text{stat})$~MeV/$c^2$, while the width is determined to be $11.4\pm1.1(\text{stat})$~MeV, where the errors are only statistical. These values are consistent with the results of the BESIII Collaboration, but the uncertainties have notably decreased. Moreover, Figure~\ref{fig2} (f) displays the joint confidence regions for the mass and width resulting from the simultaneous fit. The smallest solid ellipse corresponds to the region at the $1\sigma$ confidence level for the mass and width, while the dashed lines, ranging from inner to outer, represent the regions associated with confidence levels of $2\sigma$ to $6\sigma$, respectively.

\begin{table}[htbp] 
\caption{The mass resolutions, the mass and width of $f_{0}(980)$ reported by the BESIIII Collaboration. The bottom line is the results from the simultaneous fit, where the errors are only statistical.  \label{tab1}}
%\newcolumntype{C}{>{\centering\arraybackslash}X}
\begin{tabular}{cccc}
\toprule
\textbf{Decay Channels}	                                    & \textbf{Resolution~(MeV)}        &\textbf{Mass~(MeV/$c^2$)} &\textbf{Width~(MeV)}  \\
\colrule
\footnotesize{$\eta(1405)\to\pip\pim\piz$\cite{WuZ2012}}    &             $3.3\pm 0.2$         &       $989.9\pm0.4$      &      $9.5\pm1.1$     \\
\footnotesize{$\eta(1405)\to\piz\piz\piz$\cite{WuZ2012}}    &             $10.1\pm0.5$         &       $987.0\pm1.4$      &      $4.6\pm5.1$     \\
\footnotesize{$f_{1}(1285)\to\pip\pim\piz$\cite{XiaLG2015}} &             $3.5\pm 0.1$         &       $989.0\pm1.4$      &      $15.4\pm4.9$    \\
\footnotesize{$f_{1}(1285)\to\piz\piz\piz$\cite{XiaLG2015}} &             $9.1\pm 0.2$         &       $995.2\pm4.9$      &      $15.5\pm14.6$   \\
\footnotesize{$\chi_{c1}\to\pip\pim\piz$\cite{YanWC2018}}   &             $3.5\pm 0.1$         &       $989.8\pm1.4$      &      $10.0\pm4.0$    \\
\textbf{Simultaneous fit}                                   &                  ---             &   \bm{$990.0\pm0.4$}     &  \bm{$11.4\pm1.1$}   \\
\botrule
\end{tabular}
\end{table}

The systematic uncertainties of the mass and width of $f_{0}(980)$ arise from the fitting procedure, including the $f_{0}(980)$ signal shape, detector resolution, background shapes, and fitting ranges. To estimate the systematic errors related to the $f_{0}(980)$ signal shape, we substituted the original non-relativistic Breit-Wigner formula with the relativistic Breit-Wigner formula. Consequently, the obtained mass and width of $f_{0}(980)$ are almost unchanged, leading to relative errors of $0.004\%$ for the mass and $0.141\%$ for the width attributable to the signal shape. Regarding the systematic uncertainty originating from detector resolutions, we conducted data refitting by simultaneously increasing and decreasing the detector resolution by $1\sigma$. To be conservative, we adopted the larger difference, resulting in uncertainties of $0.0002\%$ for the mass and $3.50\%$ for the width as the final estimates for the effects of detector resolution. In order to evaluate the systematic uncertainties associated with the background shape, we randomly increased or decreased the original polynomial by one degree and performed the fitting again. As a result, we observed the maximum differences of $0.0031\%$ for the mass and $5.31\%$ for the width of $f_{0}(980)$. For the systematic uncertainty due to the fitting range, we randomly increase or decrease the fitting range of one decay process by $10-30$~MeV, and then refit the data to obtain the maximum difference, $0.0014\%$ for the mass and $4.31\%$ for the width, as the systematic uncertainties due to the fitting range. Ultimately, the total systematic uncertainties were obtained by taking the quadratic sum of the individual uncertainties. 

Consequently, we determined the mass uncertainty to be $0.0052\%$ and the width uncertainty to be $7.68\%$, respectively. After accounting for the systematic uncertainties, the mass and width of $f_{0}(980)$ were found to be $990.0\pm0.4(\text{stat})\pm0.1(\text{syst})$~MeV/$c^2$ and $11.4\pm1.1(\text{stat})\pm0.9(\text{syst})$~MeV, respectively. Here, the ﬁrst errors are statistical and the second are systematic.

%%%%%%%%%%%%%%%%%%%%%%%%%%%%%%%%%%%%%%%%%%
\section{Determination of coupling constants $g_{f\pi\pi}$ and $g_{fK\overline{K}}$ \label{coupconst}}

Because the $\fz$ can decay to both $\pi\pi$ and $KK$ final states, it can be described by the Flatt\'{e} form of the propagator~\cite{WuJJ2008,WuJJ2012}:
\begin{linenomath}
\begin{equation}
G_{f} = \frac{1}{s - m^{2}_{f} + i\sqrt{s}[\Gamma_{f\pi\pi}(s) + \Gamma_{fK\overline{K}}(s)]},
\label{Flatte}
\end{equation}
\end{linenomath}
where, $s$ is the square of $\pi\pi$ invariant mass, $m^{2}_{f}$ is the square of the nominal mass of $f_{0}(980)$ in PDG~\cite{PDG2022}, $\Gamma_{f\pi\pi}(s)$ and $\Gamma_{fK\overline{K}}(s)$ are energy-dependent partial widths of $\fz\to\pi\pi$ and $\fz\to K\overline{K}$, respectively. They are defined as:
\begin{linenomath}
\begin{eqnarray}
\Gamma_{f\pi\pi}(s)        &=& \frac{g^{2}_{f\pi\pi}}{16\pi\sqrt{s}}[\rho(\piz,\piz) + 2\rho(\pip,\pim)],  \\
\Gamma_{fK\overline{K}}(s) &=& \frac{g^{2}_{fK\overline{K}}}{16\pi\sqrt{s}}[\rho(K^{0},\bar{K}^{0}) + \rho(K^{+},K^{-})].
\label{decwidth}
\end{eqnarray}
\end{linenomath}
where, $g_{f\pi\pi}$ and $g_{fK\overline{K}}$ are coupling constants of $f_{0}(980)\to K\overline{K}$ and $f_{0}(980)\to\pi\pi$, and $\rho(A,B)=\sqrt{(s-(m_{A}+m_{B})^{2})(s-(m_{A}-m_{B})^{2})}/2s$ is the momentum of the particle A or B in the center-of-mass frame of the two-body decay. 

To determine the coupling constants $g_{f\pi\pi}$ and $g_{fK\overline{K}}$, we conducted a simultaneous fit on the $\pi\pi$ invariant mass spectra of the five decay channels. In this new fit, we replaced only the shape of the $f_0(980)$ meson with the Flatt\'{e} formula from the Breit-Wigner function, while keeping the remaining components unchanged from the previous fit. The resulting simultaneous fit provided the values for the two coupling constants $g_{f\pi\pi}=0.46\pm0.03$ and $g_{fK\overline{K}}=1.24\pm0.32$, respectively, as listed in Table~\ref{tab2}.

%%%%%%%%%%%%%%%%%%%%%%%%%%%%%%%%%%%%%%%%%%%%%%%%%%%%%%%%%%%%
\section{Joint confidence regions of the coupling constants $g_{f\pi\pi}$ and $g_{fK\overline{K}}$}
\label{confiregion}

The $f_{0}(980)$ meson is the subject of theoretical investigations that extend beyond its conventional interpretation as a quark-antiquark ($q\bar{q}$) meson~\cite{Achasov1989}. Various alternative scenarios have been proposed, suggesting its internal structure to be a tetraquark ($q^{2}\bar{q}^{2}$) state~\cite{Achasov1989,qqqq1983}, a $K\overline{K}$ molecule~\cite{KK1990}, or a quark-antiquark gluon ($q\bar{q}g$) hybrid~\cite{qqg}. The central mass of the $f_{0}(980)$ and the corresponding coupling constants $g_{f\pi\pi}$ and $g_{fK\overline{K}}$ for each of the aforementioned theoretical models were calculated and summarized in Ref.~\cite{WuJJ2008} by Jia-Jun Wu {\it et al.}. These parameters are also provided in Table~\ref{tab2} to facilitate direct comparison with the fitting results.

\begin{table}[htbp] 
\caption{The central mass and coupling constants calculated and summarized in Ref.~\cite{WuJJ2008} by Jia-Jun Wu {\it et al.} for the aforementioned theoretical predictions and the values from the simultaneous fit. \label{tab2}}
%\newcolumntype{C}{>{\centering\arraybackslash}X}
\begin{tabular}{cccc}
\toprule
\textbf{Models}	    &  \textbf{Mass~(MeV/$c^2$)} &\textbf{$g_{f\pi\pi}$~(GeV)} & \textbf{$g_{fK\bar{K}}$~(GeV)} \\
\colrule
$q\bar{q}$          &         975               &             0.64            &          1.80              \\
$q^{2}\bar{q}^{2}$  &         975               &             1.90            &          5.37              \\
$K\bar{K}$          &         980               &             0.65            &          2.74              \\
$q\bar{q}g$         &         975               &             1.54            &          1.70              \\
\textbf{Simultaneous fit} & \bm{$990.0\pm0.4$}  &    \bm{ $0.46\pm0.03$ }     &   \bm{ $1.24\pm0.32$ }      \\
\botrule
\end{tabular}
\end{table}

\begin{figure}[htbp]
\includegraphics[width=10.5 cm]{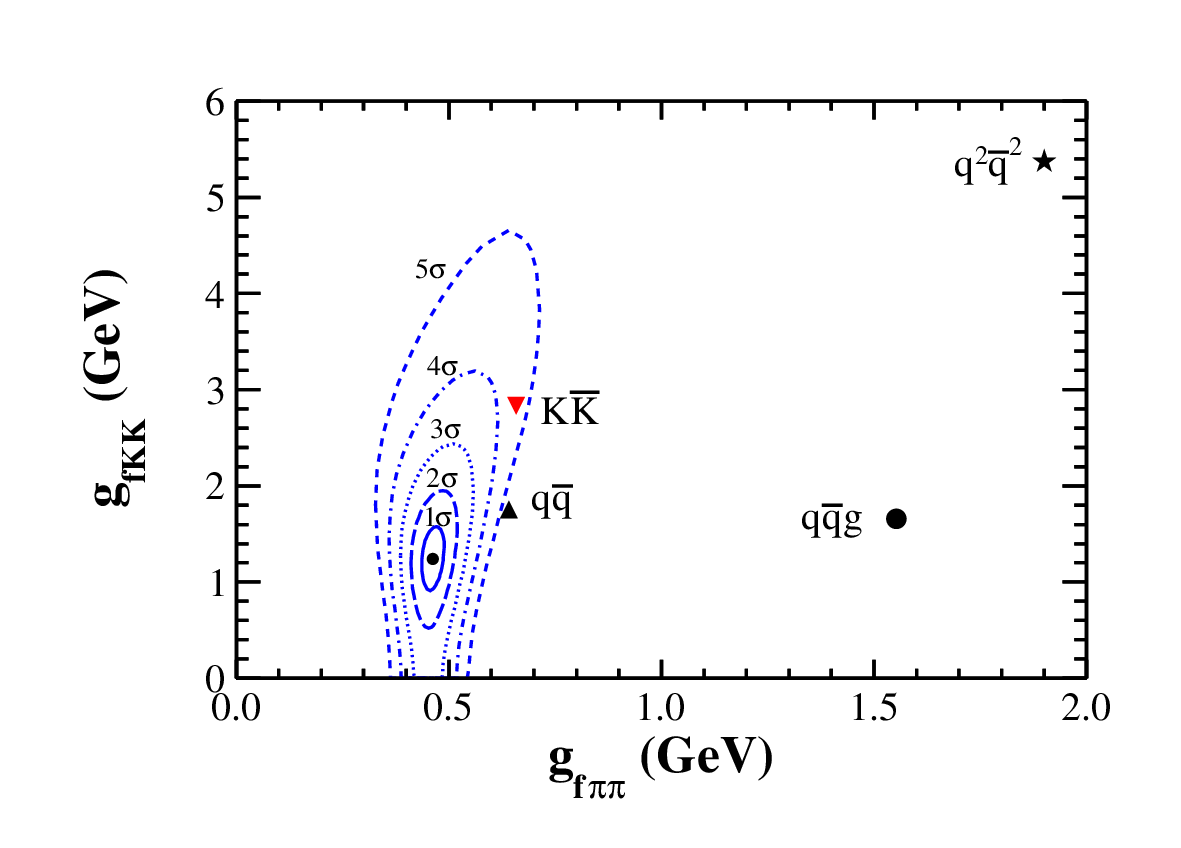}
\caption{The joint confidence regions of the two coupling constants $g_{f\pi\pi}$ and $g_{fK\overline{K}}$ from the simultaneous fit. The dashed lines from inner to outer depict the confidence levels of the two coupling constants in the order of $1\sigma$ to $5\sigma$. The triangle represents the position of the traditional quark-antiquark model, the reversed triangle represents the position of the $K\overline{K}$ molecule model, the round dot indicates the position of the quark-antiquark gluon hybrid, and the pentagon represents the position of the tetraquark model. \label{fig3}}
\end{figure}

In order to distinguish the aforementioned theoretical models based on the experimental data, we also plot the joint confidence regions of the two coupling constants resulting from the simultaneous fit, as depicted in Figure~\ref{fig3}. In the figure, the small dot represents the best estimates for the two coupling constants obtained from the fit. The smallest solid ellipse corresponds to the region at the $1\sigma$ confidence level, while the dashed lines, ranging from inner to outer, represent the regions associated with confidence levels of $2\sigma$ to $5\sigma$, respectively.

The predicted values for the two coupling constants of various theoretical models are also marked on the same plot. We observed that only the predicted position of the reversed triangle representing the $K\overline{K}$ molecule model falls within the $5\sigma$ region, whereas the predictions of other theoretical models lie outside the $5\sigma$ region. The triangle representing the quark-antiquark ($q\bar{q}$) model is positioned very close to the boundary of the $5\sigma$ region. On the other hand, the predicted positions of the tetraquarks ($q^{2}\bar{q}^{2}$) and the quark-antiquark gluon ($q\bar{q}g$) hybrid models are significantly distant from the $5\sigma$ region. Consequently, the experimental data provide support for the $K\overline{K}$ molecule model and the quark-antiquark ($q\bar{q}$) model, while tending to reject the tetraquarks ($q^{2}\bar{q}^{2}$) model and the quark-antiquark gluon ($q\bar{q}g$) hybrid model. This information proves valuable in understanding and elucidating the internal structure of the $f_{0}(980)$.

\unskip

%%%%%%%%%%%%%%%%%%%%%%%%%%%%%%%%%%%%%%%%%%
\section{Summary \label{summ}}

In summary, we conduct a simultaneous fit to five $\pi\pi$ invariant mass distributions in isospin-symmetry-breaking decays, as reported by the BESIII Collaboration. Our analysis yields the mass and width of the $f_{0}(980)$ in isospin-symmetry-breaking decay processes as $M=990.0\pm0.4(\text{stat})\pm0.1(\text{syst})$~MeV/$c^2$ and $\Gamma=11.4\pm1.1(\text{stat})\pm0.9(\text{syst})$~MeV, respectively. The first set of errors represents the statistical uncertainties, while the second set denotes the systematic uncertainties. These results are consistent with the values reported by the BESIII experiment, but with significantly improved error estimates. Furthermore, we employ the parameterized Flatt\'{e} formula to perform a simultaneous fit to the same $\pi\pi$ invariant mass distributions, enabling us to determine the model-independent coupling constants $g_{f\pi\pi}$ and $g_{fK\overline{K}}$ of the $f_{0}(980)\to K\overline{K}$ and $f_{0}(980)\to\pi\pi$ decays. Finally, we present the joint confidence regions for these two coupling constants derived from the simultaneous fit. Based on the joint confidence regions, we conclude that the experimental data provide support for the $K\overline{K}$ molecule model and the quark-antiquark ($q\bar{q}$) model, while not supporting the tetraquarks ($q^{2}\bar{q}^{2}$) model and the quark-antiquark gluon ($q\bar{q}g$) hybrid model. This information is crucial in enhancing our understanding of the internal structure of the $f_{0}(980)$.

\begin{acknowledgments}
We thank the BESIII Collaboration for publishing the data. This work is supported in part by National Natural Science Foundation of China (NSFC)
under Contract No. 11705006, and supported by the Natural Science Foundation of Henan under Grant No. 232300421140.
\end{acknowledgments}

\bibliographystyle{plain}

\begin{thebibliography}{99}

\bibitem{PDG2022} R.~L.~Workman {\it et al.} (Particle Data Group),``Review of Particle Physics,"  Prog.\ Theor.\ Exp.\ Phys.\ {\bf 2022}, 083C01 (2022).
\bibitem{SND2000} M.~N.~Achasov {\it et al.} (SND Collaboration), ``The $\phi(1020)\to\pi^{0}\pi^{0}\gamma$ decay," Phys.\ Lett.\ B\, {\bf 485}, 349-356 (2000).
\bibitem{KLOE2000} A.~Aloisio {\it et al.} (KLOE Collaboration), ``Study of the decay $\phi\to\pi^{0}\pi^{0}\gamma$ with the KLOE detector," Phys.\ Lett.\ B\ , {\bf 537}, 21-27 (2002).
\bibitem{BESII2005} M.~Ablikim {\it et al.} (BESIII Collaboration), ``Resonances in $J/\psi\to\phi\pi^{+}\pi^{-}$ and $\phi K^{+}K^{-}$," Phys.\ Lett.\ B\ , {\bf 607}, 243–253 (2005).
\bibitem{BABAR2007} B.~Aubert {\it et al.} (BABAR Collaboration), ``$e^{+}e^{-} \to K^{+}K^{-}\pi^{+}\pi^{-}$, $K^{+}K^{-}\pi^{0}\pi^{0}$ and $K^{+}K^{-}K^{+}K^{-}$ cross sections measured with initial-state radiation," Phys.\ Rev.\ D\ , {\bf 76}, 012008 (2007).
\bibitem{Martin2011PRD} R.~Garc\'{\i}a-Mart\'{\i}n, R.~Kami\'{n}ski, J.~R.~Pel\'aez, J.~Ruiz de Elvira and F.~J.~Yndur\'ain, ``Pion-pion scattering amplitude IV. Improved analysis with once subtracted Roy-like equations up to 1100 MeV," Phys.\ Rev.\ D\ , {\bf 83}, 074004 (2011).
\bibitem{Martin2011PRL} R.~Garc\'{\i}a-Mart\'{\i}n, R.~Kami\'{n}ski, J.~R.~Pel\'aez, and J.~Ruiz de Elvira, ``Precise Determination of the ${f}_{0}(600)$ and ${f}_{0}(980)$ Pole Parameters from a Dispersive Data Analysis," Phys.\ Rev.\ Lett.\  , {\bf 107}, 072001 (2011).
\bibitem{PDG2014} K.~A.~Olive {\it et al.} (Particle Data Group), ``Review of Particle Physics,"  Chin.\ Phys.\ C, {\bf 38}, 090001 (2014).
\bibitem{WuZ2012} M.~Ablikim {\it et al.} (BESIII Collaboration), ``First Observation of $\ensuremath{\eta}(1405)$ Decays into ${f}_{0}(980){\ensuremath{\pi}}^{0}$," Phys.\ Rev.\ Lett., {\bf 108}, 182001 (2012).

\bibitem{XiaLG2015} M.~Ablikim {\it et al.} (BESIII Collaboration), ``Observation of the isospin-violating decay $J/\ensuremath{\psi}\ensuremath{\rightarrow}\ensuremath{\phi}{\ensuremath{\pi}}^{0}{f}_{0}(980)$," Phys.\ Rev.\ D, {\bf 92}, 012007 (2015).

\bibitem{YanWC2018} M.~Ablikim {\it et al.} (BESIII Collaboration), ``Observation of ${a}_{0}^{0}(980)\text{\ensuremath{-}}{f}_{0}(980)$ Mixing," Phys.\ Rev.\ Lett., {\bf 121}, 022001 (2018).

\bibitem{Achasov1989} N.~N.~Achasov and V.~N.~Ivanchenko, ``On a Search for Four Quark States in Radiative Decays of $\phi$ Meson," Nucl.\ Phys.\ B, {\bf 315}, 465 (1989).
\bibitem{qqqq1983} J.~Weinstein and N.~Isgur, ``$\mathrm{qq}\stackrel{-}{\mathrm{qq}}$ system in a potential model," Phys.\ Rev.\ D\, {\bf 27}, 588 (1983).
\bibitem{KK1990} J.~Weinstein and N.~Isgur, ``$K\overline{K}$ molecules," Phys.\ Rev.\ D, {\bf 41}, 2236 (1990).

\bibitem{qqg} S.~Ishida {\it et al.}, In Proceedings of the 6th International Conference on Hadron Spectroscopy, Manchester, United Kingdom, 10th-14th July 1995 (World Scientific, Singapore, 1995), p.454.

\bibitem{Achasov1979} N.~N.~Achasov, S.~A.~Devyanin, and G.~N.~Shestakov, ``$S^{*}$-$\delta^{0}$ mixing as a threshold phenomenon," Phys.\ Lett.\ B , {\bf 88}, 367-371 (1979).

\bibitem{Kerbikov} B.~Kerbikov and F.~Tabakin, ``Mixing of the ${f}_{0}$ and ${a}_{0}$ scalar mesons in threshold photoproduction," Phys.\ Rev.\ C, {\bf 62}, 064601 (2000).

\bibitem{Achasov20041} N.~N.~Achasov and G.~N.~Shestakov, ``Proposed Search for ${a}_{0}^{0}(980)\ensuremath{-}{f}_{0}(980)$ Mixing in Polarization Phenomena," Phys.\ Rev.\ Lett., {\bf 92}, 182001 (2004).

\bibitem{Achasov20042} N.~N.~Achasov and G.~N.~Shestakov, ``Manifestation of the ${a}_{0}^{0}(980)\ensuremath{-}{f}_{0}(980)$ mixing in the reaction ${\ensuremath{\pi}}^{\ensuremath{-}}p\ensuremath{\rightarrow}\ensuremath{\eta}{\ensuremath{\pi}}^{0}n$ on a polarized target," Phys.\ Rev.\ D, {\bf 70}, 074015 (2004).

\bibitem{Kudr1} A.~E.~Kudryavtsev and V.~E.~Tarasov, ``On the possibility of observation of $a_{0}$-$f_{0}$ mixing in the $p n \to d a_{0}$ reaction," JETP\ Lett.\ , {\bf 72}, 410 (2000).

\bibitem{Kudr2} A.~E.~Kudryavtsev, V.~E.~Tarasov, J.~Haidenbauer, C.~Hanhart, and J.~Speth, ``Angular asymmetries in the reactions $ pp \to d\pi^{+} \eta$ and $ pn \to d\pi^{0}\eta$ and $a_{0}$-$f_{0}$ mixing," Phys.\ At.\ Nucl., {\bf 66}, 1946 (2003).

\bibitem{Kudr3} A.~E.~Kudryavtsev, V.~E.~Tarasov, J.~Haidenbauer, C.~Hanhart, and J.~Speth, ``Aspects of ${a}_{0}$-${f}_{0}$ mixing in the reaction $pn\rightarrow d a_{0} $," Phys.\ Rev.\ C, {\bf 66}, 015207 (2002).

\bibitem{Grishina} V.~Y.~Grishina, L.~A.~Kondratyuk, M. B\"uscher, W. Cassing, and H. Str\"oher, ``$a_{0}(980)$-$f_{0}(980)$ mixing and isospin violation in the reactions $p N\to d a_0$, $p d\to ^{3}He/^{3}H a_0$ and $dd\to ^{4}He a_0$," Phys.\ Lett.\ B, {\bf 521}, 217-224 (2001).

\bibitem{Close1} F.~E.~Close and A.~Kirk, ``Isospin breaking exposed in $f_{0}(980)$-$a_{0}(980)$ mixing," Phys.\ Lett.\ B, {\bf 489}, 24-28 (2000).

\bibitem{Close2} F.~E.~Close and A.~Kirk, ``Large isospin mixing in $\phi$ radiative decay and the spatial size of the $f_{0}(980)$-$a_{0}(980)$ mesons," Phys.\ Lett.\ B, {\bf 515}, 13-16 (2001).

\bibitem{Hanhart} C.~Hanhart, B.~Kubis, and J.~R.~Pelaez, ``Investigation of ${a}_{0}$--${f}_{0}$ mixing," Phys.\ Rev.\ D, {\bf 76}, 074028 (2007).

\bibitem{WuJJ2007} J.~J.~Wu, Q.~Zhao and B.~S.~Zou, ``Possibility of measuring ${a}_{0}^{0}(980)$-${f}_{0}(980)$ mixing from $J/\psi\rightarrow{\phi}{a}_{0}^{0}(980)$," Phys.\ Rev.\ D, {\bf 75}, 114012 (2007).

\bibitem{WuJJ2008} J.~J.~Wu and B.~S.~Zou, ``Study of ${a}_{0}^{0}(980)-{f}_{0}(980)$ mixing from ${a}_{0}^{0}(980)\rightarrow{f}_{0}(980)$ transition," Phys.\ Rev.\ D, {\bf 78}, 074017 (2008).

\bibitem{WangYD2011} M.~Ablikim {\it et al.} (BESIII Collaboration), ``Study of ${a}_{0}^{0}(980)-{f}_{0}(980)$ mixing," Phys.\ Rev.\ D, {\bf 83}, 032003 (2011).

\bibitem{WuJJ2012} J.-J.~Wu, X.-H.~Liu, Q.~Zhao and B.-S.~Zou, ``Puzzle of Anomalously Large Isospin Violations in $\ensuremath{\eta}(1405/1475)\ensuremath{\rightarrow}3\ensuremath{\pi}$," Phys.\ Rev.\ Lett., {\bf 108}, 081803 (2012).

\bibitem{Aceti2012} F.~Aceti, W.~H.~Liang, E.~Oset, J.~J.~Wu and B.~S.~Zou, ``Isospin breaking and ${f}_{0}(980)$-${a}_{0}(980)$ mixing in the $\ensuremath{\eta}(1405)\ensuremath{\rightarrow}{\ensuremath{\pi}}^{0}{f}_{0}(980)$ reaction," Phys.\ Rev.\ D, {\bf 86}, 114007 (2012).


\bibitem{DuMC2019} Meng-Chuan Du and Qiang Zhao, ``Internal particle width effects on the triangle singularity mechanism in the study of the $\ensuremath{\eta}(1405)$ and $\ensuremath{\eta}(1475)$ puzzle," Phys.\ Rev.\ D, {\bf 100}, 036005 (2019).

\bibitem{DuMC2022} Meng-Chuan Du, Yin Cheng, and Qiang Zhao, ``Vertex corrections due to the triangle singularity mechanism in the light axial vector meson couplings to ${K}^{*}\overline{K}+$c.c.," Phys.\ Rev.\ D, {\bf 106}, 054019 (2022).

\bibitem{Wang2017} En Wang, Ju-Jun Xie, Wei-Hong Liang, Feng-Kun Guo, and Eulogio Oset, ``Role of a triangle singularity in the $\gamma p\rightarrow K^{+} \Lambda(1405)$ reaction," Phys.\ Rev.\ C, {\bf 95}, 015205 (2017).

\bibitem{JingHJ2019} Hao-Jie Jing, Shuntaro Sakai, Feng-Kun Guo and Bing-Song Zou, ``Triangle singularities in $J/\psi\rightarrow\eta\pi^{0}\phi$ and $\pi^{0}\pi^{0}\phi$," Phys.\ Rev.\ D, {\bf 100}, 114010 (2019).

\bibitem{SaKai2020} Shuntaro Sakai, Eulogio Oset, and Feng-Kun Guo, ``Triangle singularity in the ${B}^{\ensuremath{-}}\ensuremath{\rightarrow}{K}^{\ensuremath{-}}{\ensuremath{\pi}}^{0}X(3872)$ reaction and sensitivity to the $X(3872)$ mass," Phys.\ Rev.\ D, {\bf 101}, 054030 (2020). 

\bibitem{Natsumi2021} Natsumi Ikeno, Raquel Molina, and Eulogio Oset, ``Triangle singularity mechanism for the $pp\ensuremath{\rightarrow}{\ensuremath{\pi}}^{+}d$ fusion reaction," Phys.\ Rev.\ C, {\bf 104}, 014614 (2021).

\bibitem{Feijoo20} A.~Feijoo, R.~Molina, L.~R.~Dai, Eulogio Oset, ``$\Lambda(1405)$ mediated triangle singularity in the $K^{-}d\to p\Sigma^{-}$ reaction," Eur.\ Phys.\ J., {\bf 82}, 1028 (2022). 

\bibitem{roofit} W.~Verkerke and D.~Kirkby, The RooFit toolkit for data modeling. Computing in High Energy and Nuclear Physics (CHEP03), La Jolla, California, USA, 24th-28th March 2003.[\text{ arXiv:physics/0306116v1 [physics.data-an]} 2003.]

\end{thebibliography}

\end{document}